\documentclass[5p,sort&compress]{elsarticle}
\usepackage[utf8]{inputenc}
\usepackage{cmap}
\usepackage[T1]{fontenc}

\usepackage{graphicx}
\usepackage{amsmath}
\usepackage{amsthm, amssymb, color}
\usepackage{sistyle}
\usepackage{calc}
\usepackage{tikz}
\usepackage{hyperref}
\hypersetup{pdfhighlight=/N,pdfborder={0 0 0}}

\newcommand{\lam}{\lambda}
\newcommand{\sig}{\sigma}

\newcommand{\skal}[1]{\left\langle#1\right\rangle}

\newcommand\subref[1]{??}

\usepackage[acronym]{glossaries}
\newacronym{kpz}{KPZ}{Kardar--Parisi--Zhang}
\newacronym{ew}{EW}{Edwards--Wilkinson}
\newacronym{rsos}{RSOS}{restricted solid-on-solid model}
\newacronym{bd}{BD}{ballistic deposition model}
\newacronym{ms}{MS}{multi-surface coding}
\newacronym{fdr}{FDR}{fluctuation-dissipation relation}
\newacronym{kk}{KK}{Kim--Kosterlitz}
\newacronym{asep}{ASEP}{asymmetric simple exclusion process}
\newacronym{tasep}{TASEP}{totally asymmetric exclusion process}
\newacronym{mbe}{MBE}{molecular beam epitaxy}
\newacronym{ks}{KS}{Kuramoto--Shivashinsky}
\newacronym{lsi}{LSI}{local scale-invariance}
\newacronym{goe}{GOE}{Gaussian orthogonal ensemble}
\newacronym{gue}{GUE}{Gaussian unitary ensemble}
\newacronym{dlg}{DLG}{driven lattice gases}

\newacronym{rng}{RNG}{random number generator}
\newacronym{lcg}{LCG}{linear congruential generator}
\newacronym{gpu}{GPU}{graphics processing unit}
\newacronym{gpgpu}{GPGPU}{general purpose computing on graphics processing units}
\newacronym{cpu}{CPU}{central processing unit}
\newacronym{simt}{SIMT}{single instruction multiple thread}
\newacronym{ram}{RAM}{computer's main memory}
\newacronym{l1}{L1 cache}{level-1-cache}
\newacronym{l3}{L3 cache}{level-3-cache}
\newacronym{lsb}{LSB}{least significant bit}
\newacronym{cu}{VP}{vector processor}
\newacronym{tdp}{TDP}{thermal design power}
\newacronym{simd}{SIMD}{single instruction multiple data}
\newacronym{mpi}{MPI}{Message Passing Interface}

\newacronym{fcc}{\texttt{fcc}}{face-centered cubic}
\newacronym{sc}{\texttt{sc}}{simple cubic}

\newacronym{cuda}{{CUDA}}{Compute unified device architecture}

\newacronym{nn}{NN}{nearest neighbor}
\newacronym{kmc}{KLMC}{3D kinetic Metropolis lattice Monte Carlo}
\newacronym{bkl}{BKL}{Bortz--Kalos--Lebowitz}
\newacronym{mc}{MC}{Monte Carlo}
\newacronym[shortplural=MCS]{mcs}{MCS}{Monte Carlo step}
\newacronym{prng}{PRNG}{pseudo random number generator}
\newacronym{lcrng}{LCRNG}{linear congruential random number generator}
\newacronym{md}{MD}{molecular dynamics}
\newacronym{pbc}{PBC}{periodic boundary conditions}
\newacronym{fbc}{FBC}{fixed boundary conditions}
\newacronym{wtm}{WTM}{waiting time method}

\newacronym{pl}{PL}{power law}
\newacronym{rhs}{r.h.s.}{right-hand side}
\newacronym{resp}{resp.}{respectively}
\newacronym{snr}{S/N}{signal-to-noise ratio}
\newacronym{fom}{FOM}{figure of merit}
\newacronym{ac}{AC}{autocorrelation}
\newacronym{su}{SU}{super-universality}

\newacronym{dd}{DD}{domain decomposition}
\newacronym{cb}{CB}{checker-board}
\newacronym{sca}{SCA}{stochastic cellular automaton}
\newacronym{rs}{RS}{random-sequential}
\newacronym{dt}{DT}{double tiling}
\newacronym{db}{DB}{dead border}
\newacronym{cdb}{cDB}{coarse dead border}
\newacronym{dtr}{DTr}{\gls{dt} \gls{dd} with random origin}
\newacronym{dtrdt}{DTrDT}{\gls{dtr} at device level and single-hit
\gls{dt} at block level}
\newacronym{dtrdb}{DTrDB}{\gls{dtr} at device level and single-hit
\gls{db} at block level}
\newacronym{dtrdtr}{DTrDTr}{\gls{dtr} at device level and single-hit
\gls{dtr} at block level}

\newacronym{tc}{TC}{thread cell}
\newacronym{t}{T}{thread}

 \makeglossaries

\begin{document}

\title{Suppressing correlations in massively parallel simulations of
lattice models}

\author[hzdrfwc]{Jeffrey~Kelling\corref{cor}}
\ead{j.kelling@hzdr.de}
\author[kfki]{G\'eza \'Odor}
\author[hzdrfwi,tuc]{Sibylle~Gemming}
\address[hzdrfwc]{
Helmholtz-Zentrum Dresden - Rossendorf,
Department of Information Services and Computing,
Bautzner Landstra\ss{}e 400, 01328 Dresden, Germany
}
\address[kfki]{
Institute of Technical Physics and Materials Science,
Centre for Energy Research of the Hungarian Academy of Sciences,
P.O.Box 49, H-1525 Budapest, Hungary
}
\address[hzdrfwi]{
Helmholtz-Zentrum Dresden - Rossendorf,
Institute of Ion Beam Physics and Materials Research,
Bautzner Landstra\ss{}e 400, 01328 Dresden, Germany
}
\address[tuc]{
Institute of Physics, TU Chemnitz,
09107 Chemnitz, Germany}
\cortext[cor]{Corresponding Author.}

\glsunset{gpu}
\glsunset{cpu}
\begin{abstract}
 For lattice Monte Carlo simulations parallelization is crucial to make studies
 of large systems and long simulation time feasible, while sequential
 simulations remain the gold-standard for correlation-free dynamics. Here,
 various \glsdesc{dd} schemes are compared, concluding with one which delivers
 virtually correlation-free simulations on \glspl{gpu}. Extensive simulations of
 the octahedron model for $2+1$ dimensional \glsdesc{kpz} surface growth, which
 is very sensitive to correlation in the site-selection dynamics, were performed
 to show self-consistency of the parallel runs and agreement with the sequential
 algorithm. We present a \gls{gpu} implementation providing a speedup of about
 $30\times$ over a parallel \gls{cpu} implementation on a single socket and
 at least $180\times$ with respect to the sequential reference.
\end{abstract}
\maketitle

\glsreset{kpz}
\glsreset{ew}
\glsreset{rs}
\glsreset{dd}
\glsreset{sca}

\section{Introduction}
 Lattice \gls{mc} simulations are employed widely to out-of-equilibrium
problems~\cite{RevModPhys.76.663}. Examples range from growth processes, such as
non-equilibrium surface growth~\cite{krug1997review} or domain growth after
phase separation~\cite{majumder2010domain}, to evolutionary game
theory~\cite{Perc2017}.  Simulations of such processes must
reproduce the physical kinetics in real systems. This is fundamentally different
from equilibrium problems, where it is admissible to change the kinetics to
speed up the relaxation, for example by cluster
algorithms~\cite{SwendsenWang1987,WOL89}. Out-of-equilibrium simulations are not
at liberty to apply such optimizations.

In practice, the most efficient way to perform lattice \gls{mc} simulations on a
bipartite lattice are checkerboard, or sub-lattice parallel~\cite{aseppardyn},
updates, which fit the definition of a \gls{sca}~\cite{Wolfram1983} since the
dynamics considers each site as strictly independent from all other sites on the
same sub-lattice. Algorithms of this type can be parallelized very efficiently
on many architectures, including
\glspl{gpu}~\cite{journals/cphysics/SchulzOON11,Weigel20123064,PaganiParisi2015,kellingOdorGemming2016_INES}.
However, in this scheme, the selection of lattice sites is correlated, which
influences the kinetics~\cite{aseppardyn}. As we have shown
recently~\cite{kellingOdorGemming2017_KPZAC}, this artificial dynamics can
indeed affect the dynamical universality class of lattice gas models.

Markov chain \gls{mc} models, like the Metropolis
algorithm~\cite{Metropolis1953} and surface growth
models~\cite{PlieschkeRacz1894,meakin,kimKosterlitz1989,odor09}, are usually
defined as a series of single-particle updates. To leave the update attempts
uncorrelated they must be performed in a \gls{rs} fashion, which cannot be
parallelized by definition. As an approximation \gls{dd} can be used, where
random site selection is restricted to a local domain per parallel
worker~\cite{PhysRevE.84.061150,KONSH2012}.

Here we review different \gls{dd} schemes for parallel \gls{gpu} viable for
implementations of lattice \gls{mc} and present one which is
virtually free of correlations. For this purpose we consider the
$2+1$--dimensional octahedron model~\cite{odor09} for \gls{kpz} surface
growth~\cite{PhysRevLett.56.889} with \gls{rs} dynamics and compare
autocorrelations for the various types of \gls{dd}.

In section~\ref{s:models} we define the octahedron model and introduce the
basics of \gls{kpz} surface growth and aging. The section concludes with a brief
summary of properties of the \gls{gpu} architecture most relevant to this work.
In section~\ref{s:DD}, the considered \gls{dd} schemes are introduced, their
impact on simulation results is presented in section~\ref{s:results},
which ends with a comparison of the performance achieved by our
implementations. We conclude in section~\ref{s:summary}.

\section{Models and methods\label{s:models}}
 \subsection{Octahedron model for ballistic deposition\label{ss:octa}}
  \begin{figure}[t]
 \centering
 \includegraphics[width=\linewidth/5*4]{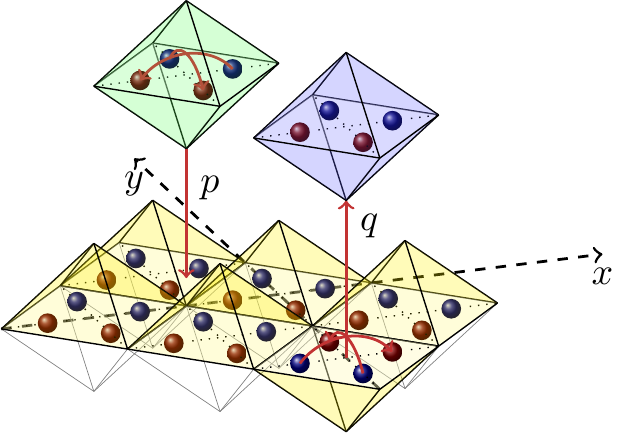}
 \caption{\label{fig:octahedron}
  Illustration of the octahedron model. The letters $p$ and $q$ label deposition
  and removal processes, respectively. The edges of the octahedra are encoded as
  up and down slopes which form a lattice gas with two species of particles drawn
  as red and blue balls, respectively. The red curved arrows connected to the
  update processes mark the directions of particle exchange in the corresponding
  lattice gas updates.
 }
\end{figure}

The octahedron model is illustrated in Fig.~\ref{fig:octahedron}: Octahedra are
deposited or removed at eligible sites with probabilities $p$ and $q$,
respectively. The slopes of the octahedra edges are mapped to a lattice gas with
the Kawasaki~\cite{PhysRev.145.224} update rules
\begin{equation}\label{eq:rule}
\left(
\begin{array}{cc}
   0 & 1 \\
   0 & 1
\end{array}
\right)
 \overset{p}{\underset{q}{\rightleftharpoons }}
\left(
\begin{array}{cc}
   1 & 0 \\
   1 & 0
\end{array}
\right)\quad,
\end{equation}
which amount to the exchange of dimers along the bisection of the Cartesian $x$
and $y$ axes. The peaks of the octahedra define the height profile of a
two-dimensional surface growing through random deposition of particles.

In the case $p=1, q=0$, which we focus on, the growth of the height is described
by the \gls{kpz} equation~\cite{PhysRevLett.56.889}:
\begin{equation}  \label{eq:kpz}
\partial_t h(\mathbf{x},t) = \sigma \nabla^2 h(\mathbf{x},t) +
\xi(\nabla h(\mathbf{x},t))^2 + \eta(\mathbf{x},t) +v \ ,
\end{equation}
where $\sigma$ models a surface tension smoothening the surface in competition
with a local growth velocity $\xi$ and an uncorrelated, zero-mean Gaussian
white noise $\eta$ roughening it. The roughness of the surface is defined as:
\begin{align}
 \label{eq:Wdef}
 W^2(L,t) &= \frac{1}{L^2} \, \sum_{i,j}^L \,h^2_{i,j}(t)  -
 \Bigl(\frac{1}{L} \, \sum_{i,j}^L \,h_{i,j}(t) \Bigr)^2\ ,
 \\
 \intertext{where $L$ denotes the lateral size of the system. It grows following
 a Family--Vicsek law~\cite{familyVicsek1985},}
 W(L,t) &\propto L^{\alpha} f(t / L^z)\ ,\nonumber
 \intertext{with the dynamical exponent $z$ and the universal growth function:}
 f(u)  &\propto
 \begin{cases}
  u^{\beta}    & \text{if } u \ll 1 \\
  {\rm const.} & \text{if } u \gg 1
 \end{cases}\label{eq:growth}
  \ .
\end{align}
The case $u\ll1$ is called the growth regime characterized by the growth
exponent $\beta$ has been determined numerically in many
studies~\cite{KellingOdorGemming2016_rsos,Halp13,PhysRevE.84.061150,FT90,kellingOdorGemming2017_KPZAC},
with the currently most precise estimate being $\beta=\num{0.2414}(2)$, which is
based on simulations using the implementation we are presenting
here~\cite{kellingOdorGemming2017_KPZAC}.

Out-of-equilibrium systems like \gls{kpz} surfaces age and can thus be best
characterized by two-time quantities like autocorrelation functions:
\begin{align}
 C(t,s) &=
 \skal{\phi(t,\mathbf{r})\phi(s,\mathbf{r})}-\skal{
 \phi(t,\mathbf{r})}\skal{\phi(s,\mathbf{r})} \ ,\nonumber
 \intertext{where $s$ denotes the waiting time, which must be in the aging regime
 $\tau_\mathrm{micro} \ll s,\ t-s \ll L^z$. One expects the scaling law:}
 C &\propto s^{-b} (t/s)^{-\lam/z}\ ,
 \label{eq:kpz_ac_C}
\end{align}
which defines the autocorrelation and aging exponents $\lambda$ and $b$,
respectively. The autocorrelation can be calculated for slope variables, i.~e.
the lattice gas:
\begin{align}
 C_s(t,s) &=
 \skal{ n(t;\vec{r}) n(s;\vec{r})} -
 {\skal{ \overline{n}(t;\vec{r})} \skal{
  \overline{n}(s;\vec{r})}}\nonumber\ ,
 \intertext{as well as the surface heights:}
 C_h(t,s) &=
 \skal{h^2(t,\mathbf{r}),h^2(s,\mathbf{r})}-\skal{
 h(t,\mathbf{r})}\skal{h(s,\mathbf{r})} \nonumber\\
&\propto s^{-b_h} \cdot (t/s)^{-\lam_\mathrm{h}/z}\ .
\intertext{For $t=s$, one finds:}
 C_h(t=s,s)
 &= W^2(L\to\infty, s) \propto s^{-b_h}\ ,
 \nonumber
 \intertext{and, after comparing to Eq.~\eqref{eq:growth}:}
  b_h = -2\beta\ , \label{eq:kpz_ac_b}
 \end{align}
which must be satisfied in the $L\to\infty$, $s\to\infty$ limit and thereby
fixes the aging exponent $b_h$ in our study.

The simulation time is measured in complete sweeps of the lattice, \gls{mcs}.

In order to characterize the finite-time behavior of quantities following
power-laws, i.~e.~$W$ and $C$, we calculate effective exponents as follows:
\begin{equation}
 e_\mathrm{eff}\left(\frac{t_i - t_{j}}2\right) = \frac{\ln E(t_i) - \ln
 E(t_{j})}{\ln(t_i) - \ln(t_{j})}\,,\label{eq:effexp}
\end{equation}
where $E\equiv W,C$; $e_\mathrm{eff} \equiv
\beta_\mathrm{eff},\lambda_\mathrm{eff}$ and $t_i > t_j$.
  \subsection{Parallel implementation\label{ss:parallel}}
  Modern computing architectures, such as \glspl{gpu}, contain large numbers of
elements executing operations in parallel. The general concept of program units
executing on these elments may be called \emph{workers}, it is those the
computational load needs to be distributed among.
In this work we apply \gls{dd} to parallelize lattice \gls{mc} algorithms. While
this confines each worker to a small region of the lattice, site selection must
remain random within this region to keep updates uncorrelated. The random memory
access patterns produced this way constitute the main obstacle for an efficient
implementation, thus we shall briefly recall the most important
concepts of \gls{gpu} architecture and its memory hierarchy, before we introduce
the \gls{dd} schemes we consider. Here, only a bird's eye view of or
CUDA~\cite{NVCPG80} implementation can be provided, which shares many technical
details with earlier implementations~\cite{KONSH2012}.

For the purposes of this work, we view \glspl{gpu} as an assembly of vector processors
or \emph{compute units} sharing \emph{global device memory} and each having its
own fast on-chip memory, which we call \emph{local memory}. This constitutes a
two-layered architecture, which \gls{dd} schemes must be designed around.
\begin{enumerate}
 \item Work must be distributed at the \emph{device-layer} among compute units.
 Since these share the global memory, we assume, that the state of the whole
 lattice is stored there.
 \item Each compute unit executes blocks of worker threads in lockstep.
 Within such a block, work must be distributed among the workers, which gives
 rise to the \emph{block-layer}. All workers in a block share access to their
 compute unit's local memory.
\end{enumerate}
In lattice models, divergence of code paths between workers can usually be
avoided easily. A more relevant problem is, that accesses of workers to the
global memory must be coalesced to be efficient, which strongly discourages
random access patterns. For this reason device-layer domains are loaded into
local memory before updates are performed. Accesses to local memory can be
random as long as bank conflicts are avoided, which can be done by padding
multidimensional data by one word.

The size of the local memory, currently \SI{48}{kB} on NVIDIA devices, imposes
constraints on the \gls{dd}. Foremost, device-layer domains must fit into this,
including any required border regions. High occupation of the device can only be
reached if a large number of threads, possibly 1024, workers are executing per
block, this means each device-layer domain must decompose into as many
block-layer domains, making the latter necessarily rather small.

Our simulations obey periodic boundary
conditions. System sizes and the length of simulation runs are chosen such that
finite-size effects are negligible.
 
\section{Domain decomposition schemes\label{s:DD}}
  \begin{figure*}[t!h]
 \centering
 \glsreset{db}
   ~\hfill
   \includegraphics[scale=1.3]{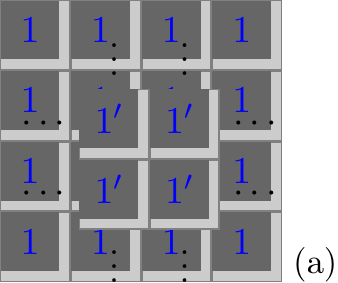}
   \hfill
 \glsreset{dt}
   \includegraphics[scale=1.3]{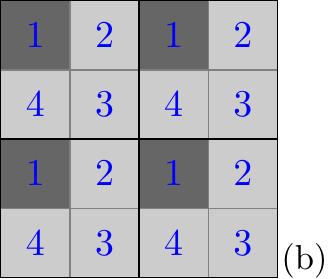}
   \hfill
 \glsunset{dt}
 \glsreset{dtr}
   \includegraphics[scale=1.3]{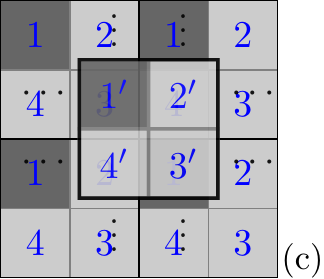}
   \hfill~
 \caption[\gls{dd} schematics]{\label{fig:dd}
  2d realizations of \gls{dd} schemes that have been evaluated for parallel
  implementations of the $(2+1)$d octahedron model. All of them could be
  straightforwardly applied in setups of arbitrary dimension. Dark areas indicate
  regions being updated concurrently while the light areas are inactive, acting
  as buffers. Numbers indicate a (randomized) sequence of synchronous steps in
  which the asynchronous domain updates are performed. Domains labeled with
  primed numbers illustrate a possible randomized decomposition after moving the
  decomposition origin: $O\to O'$. Domains calculated based on a random origin
  will always wrap around the system (\gls{pbc}), even if \gls{pbc} are not used
  in the simulation.
  \gls{db}~(a) is displayed for the special case of the octahedron
  model (and other slope-based surface growth models) where updates can affect
  neighboring cells only in positive $x$ or $y$ direction. In general, all cell
  edges would be dead borders.
 }
\end{figure*}

\gls{dd} schemes shall first be treated in a generic fashion. When the given problem
consists in distributing simulation lattice among multiple \emph{workers} which are to
perform full updates asynchronously, the solutions can be formulated independent
of the parallel architecture. The application of these schemes at device and
block layer will be detailed below.

\subsubsection{\Glsfirst{db}}
\glsreset{db}

In the \gls{db} \gls{dd} scheme, used by the \gls{gpu} implementation of the
octahedron model in~\cite{PhysRevE.84.061150,PhysRevE.89.032146}, the lattice
decomposes into tiles, where the rim of each is kept inactive during
asynchronous
updates (\emph{dead border}), so that no update of the site in the active part
of a tile would affect or be affected by the state of site in a neighboring
tile. If a border site is selected the update is not
carried out. After the asynchronous interval $t_\mathrm{async}$, the origin of
the tiling is moved randomly before the next asynchronous updates, displacing
the dead borders so that former inactive border sites may be updated and
propagate changes between tiles which were separated before. Since not all sites are
active during a sweep of the lattice, a time step under \gls{db} is a little
shorter than a whole sweep:
\begin{equation}
 1\mathrm{MCS}_{\gls{db}} = 1/({N_\mathrm{lattice sites} - N_\mathrm{border
 sites}}) \mathrm{MCS} \label{eq:dbMcs}
\end{equation}

In general, sites on the rim of tiles
in all directions need to belong to the dead border. In the octahedron model on
a square lattice, where only the links between sites carry state (the slope),
only one rim in each spatial direction needs to be inactive. This is because,
for each direction one slope is encoded on-site (to the left neighbor), hence only the
other one must be retrieved from the right neighbor.
Figure~\ref{fig:dd}(a) illustrates \gls{db} \gls{dd} in 2d.

\subsubsection{\Glsfirst{dt}}
\glsreset{dt}

Parallel implementations of lattice Metropolis Monte-Carlo
methods~\cite{Metropolis1953,newman99a} use \gls{dt} for \gls{dd}~\cite{KONSH2012}, this scheme is
illustrated in 2d in figure~\ref{fig:dd}(b). Here, the system is decomposed into
tiles, which are split into two sub-tiles in each spatial direction, creating
$2^d$ sets of non-interacting domains, where $d$ is the dimension. These domain
sets are updated in a random order, which is a permutation shuffled uniformly
at each \gls{mcs}, synchronization occurs
after completing each sweep of a domain-set. Sub-tiles do not comprise inactive
sites, thus updates of lattice sites on the rim of a sub-tile will affect sites
in neighboring sub-tiles which are at that time inactive.

In \gls{dt} borders of \gls{dd} domains always remain at the same place, which
enables higher performance, because, not having to deal with arbitrary
decomposition origins, memory alignment can be controlled and the amount of data
that needs to be exchanged between workers is
reduced. The disadvantage is that it allows errors of the effective fixed
boundary condition
approximation to accumulate which can be mediated in models with explicitly
thermally activated dynamics. This scheme is widely used for example in a
parallel implementation of the $n$-fold way
algorithm~\cite{shim2005semirigorous}.

\subsubsection{\glsfirst{dtr}}
\glsreset{dtr}

In a model where the only source of randomness is the random selection of
lattice sites, i.~e.~without explicitly thermally activated dynamics, such as
the octahedron model, site selection must not be biased. Thus, a \gls{dd} scheme which
allows imbalances in the site-selection to accumulate at specific places
(sub-tile boundaries), however slightly, cannot be expected to perform well in
general.
The accumulation can be removed by randomly moving the
decomposition origin like with \gls{db} (see figure~\ref{fig:dd}(c)), which was
also done in off-lattice simulations before~\cite{Anderson201327}. In the
\gls{dtr} scheme restricting the random origin to a coarse grid is not necessary
in any case, since there effectively are dead borders of the same width as the
decomposition domain, which easily provides enough padding to avoid having words
shared between workers.
  \subsection{Application of \gls{dd} on \glspl{gpu}\label{ss:ddGpu}}
  In a \gls{gpu} implementation \gls{dd} must be applied both at device and block
layer, where any combination of the schemes mentioned above is possible in principle.

At device layer, implementations copy the to-be-updated
decomposition domains into local memory, including all bordering region which
affect or are affected by updates. Here, thread blocks assume the role of
workers.  Synchronization occurs after a sweep of the domains has been
completed.  Communication of the boundaries is implemented by copying the data
back to global memory and loading back the new domain configuration prior to
the next sweep. It is not possible to keep non-shared parts of domains in local
memory, because it is not persistent over device synchronization events
(i.~e.~device kernel invocations).

In the implementation of the octahedron model, each 32-bit word encodes
$4\times4$ lattice sites, thus, freely picking a random origin would result in
words becoming shared between neighboring tiles. For reasons of performance, the
early implementation~\cite{PhysRevE.84.061150} restricted moving of the \gls{dd} origin in such a
way, that the borders would always be located at the edge of 32-bit words.
Essentially, the origin was moved randomly on a coarse grid with steps of
$4\times4$ lattice sites. This variation will henceforth be labeled \gls{cdb}.
\gls{db} was also implemented without restriction of the \gls{dd} origin to a coarse
grid, by increasing the width of the dead boarder to five lattice sites. This
padding ensures that if a word is shared between tiles, the left tile will not
need to write to it, because all lattice sites encoded in this word, belonging
to the left tile, will be inactive and will not have an active neighbor.

Since local memory is shared between threads, the \gls{dd} at block layer is
implemented logically, all threads access the domains assigned to them directly.
Due to the large number of threads and the limited amount of local memory, block
layer domains are rather small. The smallest size we consider here is $8\times8$
lattice sites. Performing a complete sweep of 64 updates on these would allow
fixed-boundaries errors to become too large. However, since groups of threads are
executing in lockstep and accesses to global memory do no occur while updates
are performed, synchronization of threads is cheap. This allows collective single-hit
updates to be performed, where each domain only receives one random update before
synchronization, which practically eliminates fixed-boundary effects. Here,
\gls{dt} schemes pick a domain set for each update at random, because using
a permutation would not only be computationally more expensive, but would also
decrease site-selection noise more than neceessary at this level.

When using schemes with random decomposition origin (\gls{db} or \gls{dtr}), the
origin
is moved without restrictions and thus words are likely to be shared between
neighboring \glspl{tc}. Atomic operations are therefore used to update the
domain in shared memory.

In the present work, single-hit \gls{db} is implemented using delayed borders.
This means, that updates hitting the border are not discarded, but delayed
until all bulk updates are completed. Updates to the corners of cells,
i.~e.~those which affect two borders, are carried out last.
For the octahedron model, where either $p>0$ or $q>0$, but not both,
which is the usual case when simulating the \gls{kpz} universality class,
updates can never be allowed for two \gls{nn} sites at the same time, because
slopes are restricted to $\pm 1$. Thus, in this case, delayed borders are not
required to avoid conflicts between updates and updates ignoring borders are
completely equivalent to updates with delayed borders. Hence, in the present
work, all simulations stated as using \gls{db} at block-level are actually
ignoring borders at block-level if only one of $p$ and $q$ is finite.

\subsubsection{Notation for \gls{dd} configuration}

We refer to combinations of device and block layer \gls{dd} by concatenating the
acronyms of the schemes used at both layers, putting the device layer scheme
first. The most interesting combinations discussed later are DTrDT---double
tiling with random origin at device and double tiling at block level---and
DTrDB---double tiling with random origin at device and dead border at block
level. We only consider single-hit updates at block layer.

The size and shape of thread cells is given in the notation
\texttt{TC=$\log_2(x)$,$\log_2(y)$}, where $x$ and $y$ are lateral sizes of the
cells in words (4~lattice sites). For \gls{dt} and \gls{dtr}, the smallest
possible tile size is \texttt{TC=1,1} ($8\times8$ sites), because the domains
are split in half to form sub-tiles and the implementation requires active
regions to be at least the size of one word. \gls{db} would technically allow
\texttt{TC=0,0}, but here we consider decomposition domains not smaller than
$8\times8$ lattice sites.

\section{Results\label{s:results}}
 \subsection{Correlation artifacts of \glsname{dd} schemes\label{ss:DD_oct_ac}}
  In this section we discuss the correlation artifacts of \gls{dd} schemes.
Aligned with this two-layered \gls{gpu} architecture, we will discuss
schemes at device layer first and then move on to the block layer.

\subsubsection{Device-Layer \glsname{dd}}

\glsunset{cdb}
\begin{figure}[t!]
 \centering
 \includegraphics[scale=1.3]{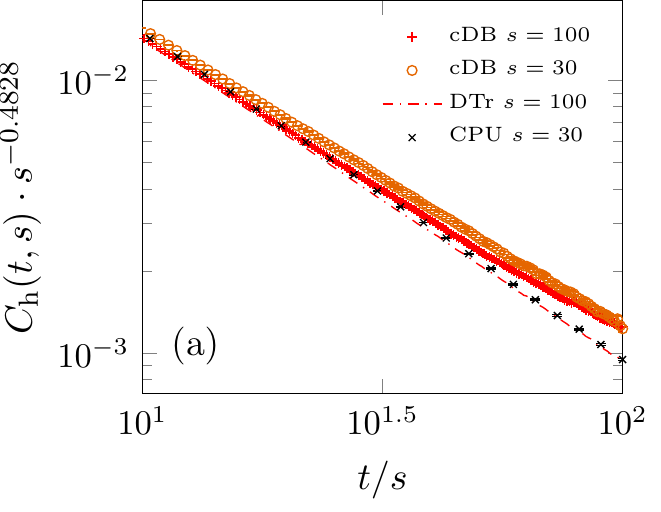}
 \\
 \includegraphics[scale=1.3]{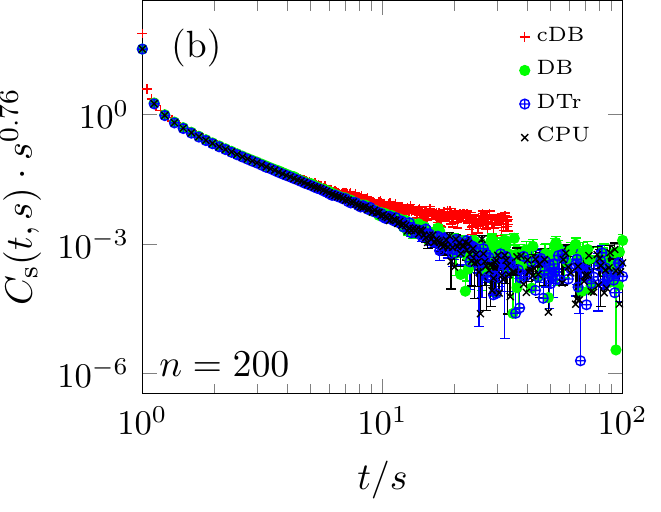}
 \caption[\gls{kpz} autocorreclations with different
 device-layer \glspl{dd} for \gls{rs}.]{\label{fig:acdlddh}
  Autocorrelations under \gls{rs} dynamics with different device-layer
  \glspl{dd}.
  (a) Autocorrelation of heights (data from~\cite{PhysRevE.89.032146}). The
  system sizes are $L=2^{15}$, $L=2^{16}$ and $L=2^{13}$ in the \gls{cdb}, the
  \gls{dtr} and the sequential (\gls{cpu}) runs, respectively. Sample sizes are
  $n_=696, 830, 71, 4367$, in the as they appear in the legend form top to
  bottom.
  (b) Autocorrelation of slopes for different types of device-layer \gls{dd}
  compared to a sequential simulation at $L=2^{13}$. All sample sizes are
  $n=200$, so noise levels can be compared visually. The presented parallel runs
  use single-hit \gls{dt} at block-layer.
 }
\end{figure}

\glsreset{cdb}
\glsreset{db}
\glsreset{dtr}
\glsreset{dt}

In an aging study published in~\cite{PhysRevE.89.032146} it was found, that the
\gls{gpu} implementation of the octahedron model with \gls{rs} dynamics used
therein 
exhibits an asymptotic autocorrelation function deviating from the
sequential \gls{cpu} reference implementation. This parallel implementation uses
\gls{cdb}, where the \gls{dd} origin is only moved on a coarse grid with $4\times4$
lattice-site units.
Figure~\ref{fig:acdlddh} compares autocorrelation functions of both heights~(a) and
slopes~(b) resulting when \gls{cdb} is employed at device-layer with other schemes
and sequential simulation results. For the slopes one clearly
observes convergence of the autocorrelation function to a finite value. The
autocorrelation of heights under \gls{cdb} can also be seen to deviate from the
\gls{pl} laid-out by the sequential reference. A finite asymptotic limit 
has not been reached in any of these simulations due to the short time-scale
considered.

Intuitively, a finite asymptotic value for the autocorrelation function
could mean that there is some pattern imprinted on the system during
its evolution. When the \gls{dd} origin is only shifted on a coarse grid, only
lattice sites which lie on edges of this coarse grid can become borders. These
sites are thus updated less frequently than the remaining sites, which never
become border-sites. Hereby, two types of sites are defined in the system evolving at
different rates. Since these borders are only a single lattice site wide and
device-layer domains are very large, the effect of this is apparently too small
to be observed in the kinetics of surface roughening or steady state properties.
It is, however, strong enough to imprint a persistent pattern onto the surface,
which can be observed in the autocorrelation functions.

\glsunset{db}
A straightforward way to solve this problem, would be to not shift the \gls{dd}
origin on a coarse grid but allow arbitrary coordinates, which does indeed
eliminate the observed correlations. For the present bit-coded implementation,
unrestricted \gls{db} requires borders which are five inactive lattice sites wide.
Moving the origin freely ensures that all lattice sites are updated with the same
frequency, when sufficiently long times are considered.
Ideally, border sites are only inactive for one asynchronous update sweep
at a time ($t_\mathrm{async}$). But, after moving the \gls{dd} origin randomly,
the old and the new borders will necessarily intersect at a grid of points,
which are then inactive for $2t_\mathrm{async}$. Due to the wide borders, these
intersections cover patches of $5\times5$ lattice sites. Thus, the wide borders
are producing locally varying update frequencies at short time scales. This
manifests as additional noise in the autocorrelation functions and possibly
other observables.

All curves presented in figure~\ref{fig:acdlddh}(b) are averaged
over the same number $n=200$ of runs. At late times ($t/s>30$) autocorrelation
signals have decayed below the noise-level of the respective sample, so the
variance can be compared. Reducing $t_\mathrm{async}$ decreases adverse effects
of \gls{dd} borders and thus also the additional noise caused by wide borders.
The data shown for \gls{db} stems from simulation using $t_\mathrm{async} =
\SI{.5}{MCS}$. Even larger noise is observed for $t_\mathrm{async} =
\SI{1}{MCS}$.

This study shows \gls{dtr} to be superior to the other presented schemes: It
appears to neither introduce correlation nor additional noise, compared with a
sequential simulation. It has only the disadvantage of using smaller active domains for
asynchronous update sweeps, but this influence is negligible for the large
domain sizes usually used at device-layer and it serves to keep update
frequencies homogeneous. For this reason it may even be preferable over
\gls{db} with a border-width of one lattice site, since intersections of thin
borders could still cause tiny imbalances. Using \gls{dt} without randomly
moving the \gls{dd} origin results in similar correlations as exhibited by
\gls{cdb}.

As the goal of this work is present a virtually correlation-free approach, all
further considerations are based on \gls{dtr} as device layer \gls{dd}.

\subsubsection{Block-Layer \glsname{dd}}

\begin{figure}[t!]
 \centering
 {
  \includegraphics[scale=1.3]{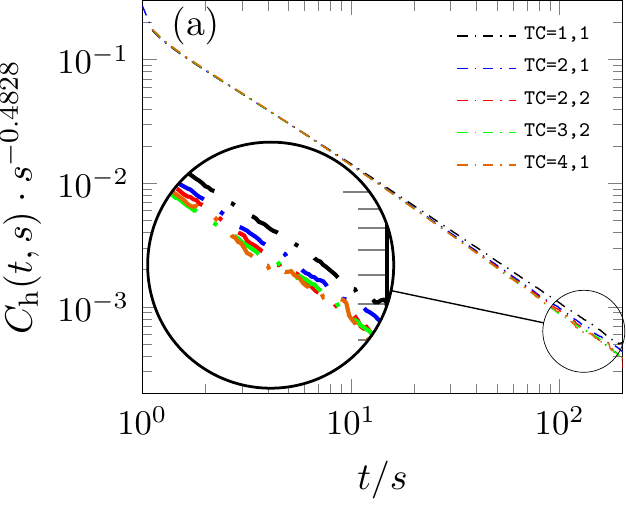}
 }
 {
 }\\
 {
 }
 {
  \includegraphics[scale=1.3]{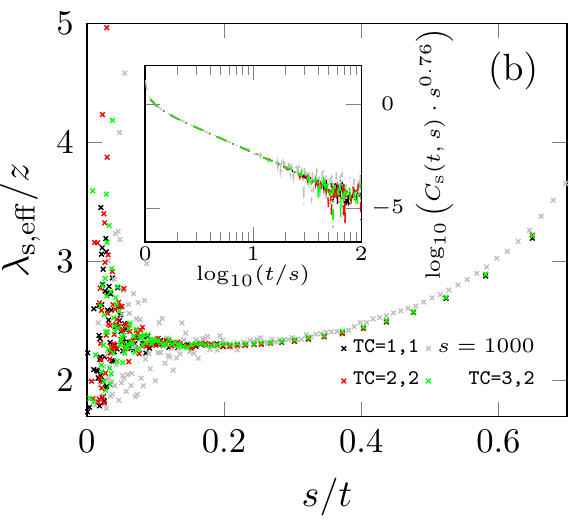}
 }
 \caption[Overview of \gls{rs} Autocorrelation results using \gls{dtrdt}.
 ]{\label{fig:acbldt}
  This figure presents the effects of different \gls{tc} configurations on
  \gls{rs} autocorrelation results when using \gls{dt} for \gls{dd} at block-layer
  (\gls{dtrdt}). (a): Comparison of different \gls{tc}
   volumes and shapes for waiting time $s=100$. (b) Comparison of
   \gls{ac} of slopes for different \gls{tc} configurations at $s=100$ and, for
   the smallest configuration \texttt{TC=1,1}, $s=1000$.\\
   The system size for
   \texttt{TC=2,1} is $L=2^{17}$, all other system sizes are $L=2^{16}$. Sample
   sizes vary: $n_{\texttt{TC=1,1}}\geq 38; n_{\texttt{TC=2,1}}\geq120;
   n_{\texttt{TC=2,2,T4,3}}\geq71; n_{\texttt{TC=2,2,T5,4}}\geq20;
   n_{\texttt{TC=3,1}}\gtrsim n_{\texttt{TC=1,3}}\geq78;
   n_{\texttt{TC=3,2}}\geq28$ and $n_{\texttt{TC=4,1}}\geq30$. Statistical
   $1\sigma$-errors are below $\num{5e-4}$ for all data points.
 }
\end{figure}

The observed changes in autocorrelation functions with block-layer \gls{dd} are
so small, that they could not be resolved by most sequential simulations. At the
same time sequential and parallel simulations may differ by small corrections,
not affecting any universal properties.
Thus, even if a small difference between
parallel and sequential results could be resolved, it would be unclear if this
was a cause for concern. For these reasons, comparisons with sequential
simulations would not be constructive at this point.

In simulations using \gls{dd}, the size and shape of domains remain free
parameters, where the exact sequential behavior corresponds to the limit of
infinite domain size. This view suggests checks for self-consistency as a viable
method for this analysis.

Figure~\ref{fig:acbldt}(a) shows autocorrelation functions for heights when
using \gls{dtrdt} for different block-layer domain configurations.
Different \gls{tc} configurations appear to show a trend in the autocorrelation
functions at late times:
\[
 C_\texttt{TC=1,1} > C_\texttt{TC=2,1} > C_\texttt{TC=2,2} \gtrsim C_\texttt{TC=3,2}
 \approx C_\texttt{TC=4,1}
\]
For the most part, the differences are barely significant, but quite clear
between the configuration with the smallest volume
block-layer domains, \texttt{TC=1,1}, and the others. If
lateral cell dimensions dominated, the configuration \texttt{TC=4,1} would
be expected to give more similar results, instead it much better agrees with
\texttt{TC=3,2}, which features block-layer domains with the same volume
($V_\texttt{TC=4,1}=V_\texttt{TC=3,2}=16\cdot2^5=256$ lattice sites).

Contrary to the these observations for the autocorrelation of heights, the
autocorrelation of slopes is identical in all these simulations. This is
shown in figure~\ref{fig:acbldt}(b) by overlaying the effective
exponents $\lam_{\mathrm{s,eff}}/z$, corresponding to the autocorrelation
functions of the slopes shown in the inset.

\begin{figure}[t!]
 \centering
 \includegraphics[scale=1.3]{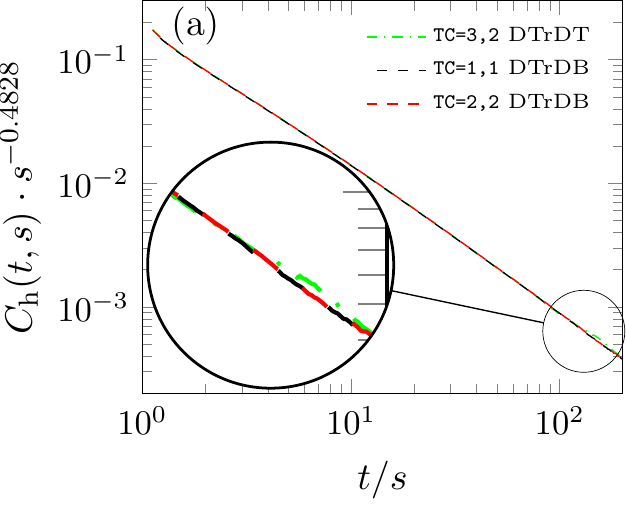}
 \\
 \includegraphics[scale=1.3]{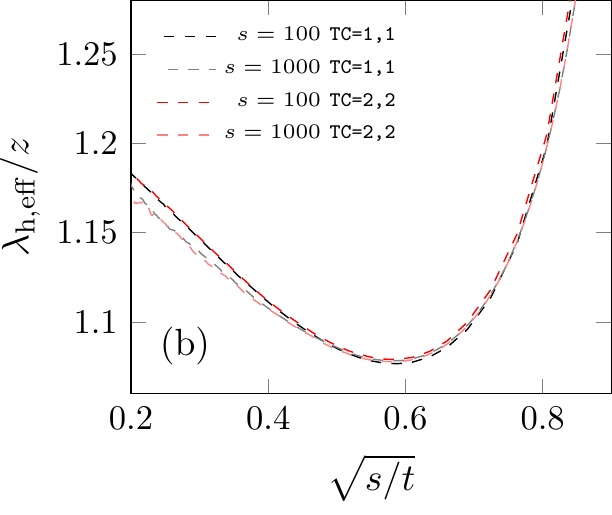}
 \caption[Overview of \gls{rs} Autocorrelation results using \gls{dtrdb}.
 ]{\label{fig:acbldb}
  Comparison of autocorrelation results in \gls{rs} simulations using \gls{db}
  for \gls{dd} at block-layer (\gls{dtrdb}). (a) Autocorrelation functions for
  different \gls{tc} sizes for waiting time $s=100$. Results using \gls{dtrdt}
  with \gls{tc} configuration \texttt{TC=3,2} are given for comparison. (b)
  Local slope analysis of \gls{dtrdb} results for $s=100$ and $1000$. All
  system sizes are $L=2^{16}$. Sample sizes are: $n_{\texttt{TC=1,1}}\geq1044;
  n_{\texttt{TC=2,2}}\geq708$ and (\gls{dtrdt}) $n_{\texttt{TC=3,2}}\geq28$.
  Samples sizes of \gls{dtrdb} runs are much larger because these stem from
  production runs which where used to extract final results.
 }
\end{figure}

The reason \gls{dtrdt} produces some small correlated noise may be found in the
grid-like site-selection pattern at block-layer, where the coordinates for
collective update attempts are restricted to the selected set of active domains,
which resembles a grid. This grid-pattern cannot be eliminated by randomly
moving the \gls{dd} origin at the block-layer after each collective update
(\glsname{dtrdtr}). We also observed, that \gls{dtrdtr} does not substantially
improve on the results obtained using \gls{dtrdt}, which is why it was not
investigated in further detail.

The grid can be eliminated by turning back to the \gls{db} scheme at block layer
(\glsname{dtrdb}).
Figure~\ref{fig:acbldb} shows autocorrelation data for surface heights from
\gls{dtrdb} simulations using \texttt{TC=1,1} and \texttt{TC=2,2}. Here, the
autocorrelation functions are in perfect agreement. Panel~(b) shows local
slope analyses for autocorrelation functions with waiting times $s=100$ and
$1000$ from these simulations, which also agree almost perfectly. The data
presented for \gls{dtrdb} is taken from production runs, which are further
analyzed in the {next} section. The curves are
much smoother because of the larger sample size afforded.

Figure~\ref{fig:acbldb}(a) also shows a curve from \gls{dtrdt}
simulations with \texttt{TC=3,2} for comparison. A good agreement cannot be
denied, suggesting, that at this block-layer domain size, the \gls{dtrdt}
simulations are sufficiently converged with respect to the autocorrelation of
heights.
  \subsection{Corrections to scaling\label{ss:DD_oct_scaling}}
  The roughness of a \gls{kpz} surface is expected to grow with a power law. The
asymptotic growth exponent in $2+1$ dimensions is only known numerically.

Eq.~\eqref{eq:growth} allows for a non-universal, constant prefactor to the
growth-law. Such model-dependent parameters are affected by \gls{dd}. One aspect
we wish to point out is, that the size of decompositon domains influences the
amplitude of the noise parameter $\eta$ in Eq.~\eqref{eq:kpz}, due to the
variance of the rate at which each site is selected for updates decreasing with
the size of the smallest decomposition domains. In other words, the system is
sampled more smoothly with fine \gls{dd}, which slightly inhibits the roughening
of the surface, even when updates are uncorrelated, leading to a smaller
prefactor to the growth law for finer \gls{dd}. As a consequence, data from
simulations with different \gls{dd} cannot be averaged. Furthermore, in studies
where absolute values of the roughness are relevant, such as finite-size scaling,
the coarseness of the \gls{dd} must be kept fixed.

\begin{figure}[t!]
 \centering
 \includegraphics[scale=1.3]{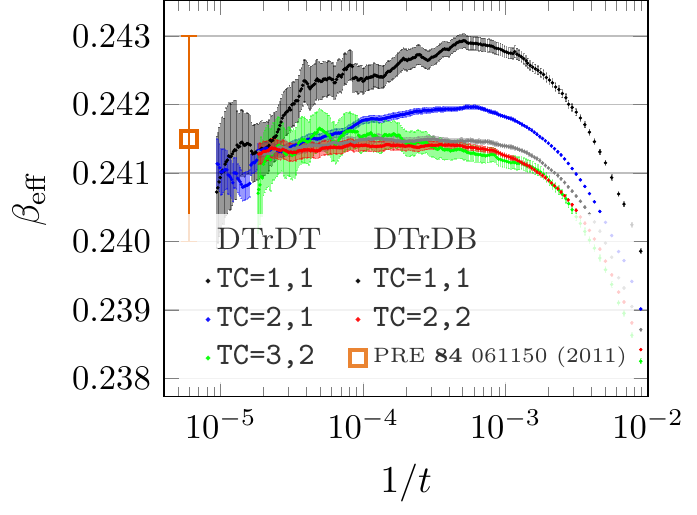}
 \caption[\gls{kpz} state roughness growth under \gls{rs}
 dynamics]{\label{fig:scalingWidthRs}\label{fig:kpzBeta}
 Effective exponents for selected \gls{dd} configurations. System sizes
 are $L_{\text{\gls{dtrdt}},\texttt{TC=2,1}}=2^{17}$ others $L=2^{16}$. Sample sizes
 are: $n_{\text{\gls{dtrdt}},\texttt{TC=1,1}}\geq85$,
 $n_{\text{\gls{dtrdt}},\texttt{TC=2,1}}\geq396$,
 $n_{\text{\gls{dtrdt}},\texttt{TC=3,2}}\geq89$,
 $n_{\text{\gls{dtrdb}},\texttt{TC=1,1}}\geq1107$ and
 $n_{\text{\gls{dtrdb}},\texttt{TC=2,2}}\geq708$.
 Propagated $1\sig$ error bars are attached to the effective exponents.
 The orange square shows the asymptotic exponent $\beta=\num{.2415}(15)$
 obtained in Ref.~\cite{PhysRevE.84.061150}.
 }
\end{figure}

Figure~\ref{fig:scalingWidthRs} shows effective scaling exponents for a selected
set of \gls{dd} configurations. All curves suggest about the same asymptotic
value $\beta$, but they differ at finite times, i.~e.~in corrections to scaling.
Because only the asymptotic scaling exponent $\beta$ is a universal property,
while $\beta_\mathrm{eff}(t)$ depends on the model for finite times, this does
not mark the result of any of these simulations wrong. However, the effective
exponents from \gls{dtrdb} runs show a plateau over almost two decades,
suggesting that corrections at late times are very small.

The figure also shows the asymptotic value $\beta=\num{.2415}(15)$ from
Ref.~\cite{PhysRevE.84.061150}, which is based on \gls{gpu} simulations using
\gls{cdb} at device and single-hit \gls{db} at block level. This value agrees well
with the data from \gls{dtr} implementations despite the presence of
correlations.
  \subsection{Performance\label{ss:ddBench}}
  \begin{figure}[t!]
 \centering
 \includegraphics[scale=1.1]{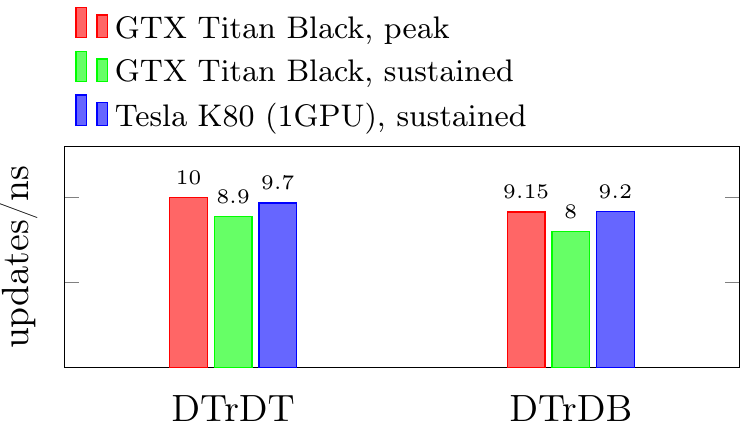}
 \caption[\glsname{rs} octahedron model benchmarks]{\label{fig:parallelRsBench}
  Performance of the \gls{dtrdt} and \gls{dtrdb} variations of the \gls{rs}
  implementation of the octahedron model on \gls{gpu}. For K80, the sustained
  performance equals the peak performance. Benchmarks were performed for systems
  of lateral size $2^{16}$.
}
\end{figure}

The performance of the two most interesting implementations for future
applications on NVIDIA Kepler-generation \glspl{gpu} is compared in
figure~\ref{fig:parallelRsBench}.  Both use \gls{dtr} at device-layer and differ
in the type of block-layer \gls{dd} employed: \gls{dt} and \gls{db}.
On a GTX Titan Black GPU, the performance drops
after the first $\sim\SI{100}{MCS}$, because the device clocks down under load.
This leads to a measured sustained performance which is lower than the peak. The
performance on a Tesla K80 is about constant. The \gls{dtrdb} variant is
consistently slower, by about ten percent.

A sequential implementation on an Intel i7-4930k \gls{cpu} delivers $\num{0.055}$
update attempts per \SI{}{ns} for a system of lateral size $L=2^{12}$, a size
where the whole system fits into the L3 cache. The performance is less for
larger systems, which do not fit into L3 cache. Running multiple independent
runs on the same device is also likely to reduce performance. A parallel
implementation, using \gls{dtr}, running twelve threads, performs $\num{0.28}$
update attempts per \SI{}{ns}. This leads to a speedup factor of about 30 for
the \gls{gpu} code over a single-socket \gls{cpu}.
 
\section{Summary and Conclusions\label{s:summary}}
 In this work we compared \gls{gpu} implementations of the octahedron model for
surface growth using different \gls{dd} approaches. We were able to identify and
eliminate the cause of correlations in a previously published \gls{gpu}
simulation~\cite{PhysRevE.89.032146}. Our solution applies the \glsfirst{dtr}
scheme at device layer, which can be expected to perform well in simulations for
any coarse decomposition with full sweeps between synchronizations. Apart from
an obvious reduction in the site-selection noise, the results are indistinguishable
from sequential runs on \gls{cpu} up to the accuracy those can provide.

To go beyond this precision, we checked for artifacts due to  fine \gls{dd} at
block layer by testing the self-consistency of \gls{gpu} simulation results. We
found, that using single-hit \glsfirst{db} at block level gives self-consistent
results for both autocorrelation and roughness growth. The differences at finite
times, we saw when using single-hit \glsfirst{dt} for this \gls{dd} layer, do
not allow a firm conclusion that universal properties could be affected.
However, it is advantageous to exclude any finite-time corrections, which are
artifacts of the simulation method applied. For these reasons we conclude, that
the combined \gls{dtrdb} scheme is the best one to be used in similar lattice
\gls{mc} studies. Our implementation of the octahedron model using this scheme
delivers about nine update attempts per $\mathrm{ns}$ on high-end
Kepler-generation NVIDIA \glspl{gpu}.

The code used in this work, together with code used
in~\cite{kellingOdorGemming2017_KPZAC,kellingOdorGemming2016_INES} can be found
at~\url{https://github.com/jkelling/CudaKpz}.
 
\vskip 1.0cm

\noindent
{\bf Acknowledgments:}\\

Support from the Hungarian research fund OTKA (Grant No.~K109577), the
Initiative and Networking Fund of the Helmholtz Association via the W2/W3
Programme \mbox{(W2/W3-026)} and the International Helmholtz Research School
NanoNet \mbox{(VH-KO-606)} is acknowledged.
We gratefully acknowledge computational
resources provided by the HZDR computing center, NIIF Hungary and the Center for
Information Services and High Performance Computing (ZIH) at TU Dresden.

 \bibliography{article}
 \bibliographystyle{unsrt}

\end{document}